\documentclass[%
 reprint,
 amsmath,amssymb,
 aps]{revtex4-2}

\usepackage{graphicx}
\usepackage{dcolumn}

\newcommand{\Mp}{M_{\rm pl}}

\newcommand{\bbm}[1]{\hbox{\boldmath{$#1$}}}
\newcommand{\sbm}[1]{\hbox{\boldmath{\scriptsize$#1$}}}
\newcommand{\dNex}{g$\delta N$ formalism}
\newcommand{\Lee}{{{\cal L}\hspace{-6pt}\raise 1.5pt \hbox{-}\,}}
\newcommand{\If}{\hat{\cal I}_{(\alpha)f}}

\begin{document}


\title{Statistical anisotropy of primordial gravitational waves from generalized 
$\delta N$ formalism}

\author{Takahiro Tanaka}
\email{t.tanaka@tap.scphys.kyoto-u.ac.jp}
\affiliation{%
Department of Physics, Kyoto University, Kyoto 606-8502, Japan
}%
\affiliation{Center for Gravitational Physics and Quantum Information, Yukawa Institute for Theoretical Physics, Kyoto University, Kyoto 606-8502, Japan}

\author{Yuko Urakawa}
\email{yukour@post.kek.jp}
\affiliation{Institute of Particle and Nuclear Studies, High Energy Accelerator Research Organization (KEK), Oho 1-1, Tsukuba 305-0801, Japan}%
\affiliation{
The Graduate University for Advanced Studies (SOKENDAI), Tsukuba 305-0801, Japan
}
\affiliation{International Center for Quantum-field Measurement Systems for Studies of the
Universe and Particles (QUP), High Energy Accelerator Research Organization
(KEK), Tsukuba, Ibaraki 305-0801, Japan}

\date{\today}

\begin{abstract}
In this letter, we demonstrate how to use the generalized $\delta N$ formalism, which enables us to compute the evolution of all the large scale fluctuations, including gravitational waves, solely by solving the evolution of the background homogeneous Universe. Using the Noether charge density, we derive an analytic formula which describes the mapping between the fluctuations at the horizon crossing and the sourced gravitational waves at the end of inflation. This formula can apply also to an inflation model with an anisotropic background. Using this formula, we discuss the condition for the non-vanishing linear polarization and the qualitative difference between single- and multi-gauge field models. 
\end{abstract}

\maketitle


{\it Introduction.}-- The $\delta N$ formalism~\cite{Starobinsky:1982ee, Sasaki:1995aw, Sasaki:1998ug}, which is based on the separate universe approach~\cite{Salopek:1990jq, Wands:2000dp}, has played the central role to connect the prediction of each inflation model with various observations, enabling a simple computation of the superhorizon dynamics and also providing an intuitive understanding on the evolution of primordial fluctuations. However, their application was limited to a system which contains only scalar fields and it could not be used even to compute gravitational waves (GWs). In Ref.~\cite{Tanaka:2021dww}, we have shown that the separate universe approach can be applied under a rather general setup. Based on this, we can generalize the $\delta N$ formalism to compute the adiabatic perturbation and GWs sourced by non-zero spin fields. In this letter, we demonstrate this computation, considering a model with $U(1)$ gauge fields which have the kinetic mixing with scalar fields, which ubiquitously appear e.g. in the 4D low energy effective field theory of string theory.

{\it \dNex.}-- 
The $\delta N$ formalism computes the evolution of fluctuations at the leading order of the gradient expansion \cite{Salopek:1990jq, Shibata:1999zs}, which is an expansion scheme with respect to the spatial gradient. The gradient expansion starts with smoothing the small scale fluctuations. As a consequence of the smoothing, operating the spatial gradient gives rise to the suppression by a small quantity $\epsilon$, which is usually characterized by the spatial variation of the fields within each causally connected patch. At the leading order of the gradient expansion, we simply send $\epsilon$ to 0. As shown in \cite{Tanaka:2021dww}, the separate universe approach and the $\delta N$ formalism generically applies to a theory which satisfies the spatial diffeomorphism invariance and the locality. 

Under these two conditions, one can compute the time evolution of the inhomogeneous Universe simply by solving a set of the corresponding ordinary differential equations for different initial conditions specified at around the horizon crossing. The conventional $\delta N$ formalism can be applied only to a system with scalar fields, while the generalized $\delta N$ formalism (\dNex) can be applied broadly to a general model that satisfies the above-mentioned two conditions \cite{Tanaka:2021dww}.

{\it Preliminaries.}-- 
In this letter, we consider a system with $D$ scalar fields $\phi^I$ and $D'$ U(1) gauge fields $A^\mu_{(\alpha)}$ whose Lagrangian density is given by
\begin{align}
    {\cal L}_{\rm mat} &= {\cal P}(X^{IJ},\, \phi^I) -  \frac{f_{(\alpha)}^2(\phi^I)}{4} F_{\mu \nu(\alpha)} F^{\mu \nu}_{(\alpha)} 
    \,,  \label{Exp:Lmatter_single}
\end{align}
where $F_{\mu \nu(\alpha)}$ denotes the field strengths of the gauge fields $A^\mu_{(\alpha)}$. We do not explicitly write the summation over the label of the gauge fields $\alpha$.

We express the 4-dimensional line element as
\begin{align}
    ds^2 = - N^2 dt^2 + g_{ij} (dx^i + N^i dt) (dx^j + N^j dt)\,, 
\end{align}
with $i, j = 1, \cdots\hspace{-2pt}, 3$. We express the spatial metric as $g_{ij} = e^{2 \psi}\, \gamma_{ij}$, where $\gamma_{ij}$ satisfies $\det[\gamma]=1$. 
Using $\psi$, the determinant of $g_{ij}$ is given by $g= e^{6\psi}$. In the separate universe evolution, we need to employ local gauge conditions~\cite{Tanaka:2021dww}. Here, we adopt $N_i=0$ and $A_{0}=0$. In $N_i=0$ gauge, the expansion $K$ and the shear ${A^i}_j$, given by the trace part and the traceless part of the extrinsic curvature, read $K = 3 \dot{\psi}/N$ and ${A^i}_j = \gamma^{im} \dot{\gamma}_{mj}/2N$, respectively, with a dot being the time derivative with respect to $t$. We determine the time slicing at the horizon crossing, requiring $\delta \psi (t_*,\, \bbm{x})=0$ and the one at $t=t_f$, requiring $\delta K (t_f,\, \bbm{x})=0$. The residual gauge degrees of freedom are eliminated by imposing additional gauge conditions on the initial condition at the horizon crossing~\cite{Tanaka:2023}. We introduce $\delta {\gamma^i}_j$ and $\delta \gamma_{ij}$ as
$\gamma_{ij}(t,\, \bbm{x}) \equiv \bar{\gamma}_{ik}(t) \left[ e^{\delta \gamma(t,\, \sbm{x})} \right]^k\!_j$ and $\delta \gamma_{ij}(t,\, \bbm{x}) \equiv  \bar{\gamma}_{ik}(t) \delta {\gamma^k}_j(t,\, \bbm{x})$. Here and hereafter, we put a bar on background variables. 

The Maxwell equations are given by $\dot{\pi}_{(\alpha)}^i = {\cal O}(\epsilon^2)$ with the conjugate momenta being
\begin{align*}
    \pi_{(\alpha)}^i \equiv \frac{\partial (N \sqrt{g} {\cal L}_{\rm matter}) }{\partial \dot{A}_{i(\alpha)}} =   \sqrt{g} f_{(\alpha)}^2(\phi^I) g^{ij} \frac{\dot{A}_{j(\alpha)}}{N}  \,, 
\end{align*}
which correspond to the Noether charge densities at the leading order of the gradient expansion~\cite{Tanaka:2021dww}. The energy densities of the gauge fields are given by
\begin{align}
    \rho_{A(\alpha)} =  \frac{\gamma_{ij} \pi_{(\alpha)}^i \pi_{(\alpha)}^j}{2 f_{(\alpha)}^2 e^{4 \psi}}  + {\cal O}(\epsilon^2)\,,  \label{Exp:rhoA}
\end{align}
which indicates that $\rho_{A(\alpha)}$ grows, when $f_{(\alpha)}$ decreases faster than $e^{-2 \psi}$.

{\it Primordial GWs.}-- In this letter, we assume the near de Sitter evolution and that the shear is much smaller than the expansion, i.e., 
\begin{align}
  ({\rm A1})\quad  - \frac{3 \dot{K}}{NK^2} \ll 1 \,, \quad  \frac{{A^i}_j {A^j}_i}{K^2} \ll 1\,. \label{Cond:smallshear}
\end{align}The second assumption in (A1) can be verified when the summation of the energy densities of all the gauge fields remains much smaller than the total one, $\rho$. From the conservation of the Noether charge density~\cite{Tanaka:2021dww}, we can obtain $\gamma_{ij}$ evaluated at the reheating surface $t_f$ as (the derivation can be found in Supplementary material)  
\begin{align}
   & \gamma_{ij} (t_f,\, \bbm{x}) = \gamma_{ij*}(\bbm{x}) 
    -  2 \left[ \gamma_{il*}(\bbm{x}) \gamma_{jm*}(\bbm{x}) \right]^{\rm TL} \nonumber \\
    & \qquad \qquad \quad \times \pi_{(\alpha)}^l(\bbm{x}) \pi_{(\alpha)}^m(\bbm{x}) \, {\cal I}_{(\alpha)f}(\{\varphi^{a'}_*\}')+ {\cal O}(\epsilon^2) \,, \label{Sol:gamma}
\end{align}
with ${\cal I}_{(\alpha)f}$ being
\begin{align}
    & {\cal I}_{(\alpha)f}(\{\varphi^{a'}_*\}') \simeq \frac{1}{\rho_{*}} \int^{\psi_f}_{\psi_*}  \frac{d \psi}{f^2_{(\alpha)}(\phi^I) e^{4 \psi}} ,  \label{Def:calIwohat} 
\end{align}
where $\{\varphi^{a'}_*\}'$ denotes the set of the fields by which we provide the initial condition at the horizon crossing $t_*$ (a detailed argument can be found in \cite{Tanaka:2023}) and $\rho_*$ denotes the total energy density at $t=t_*$. Here, TL denotes the traceless part about the $(i, j)$ indices defined by using the spatial metric $\gamma_{ij*}$. In Eq.~(\ref{Sol:gamma}), only the leading contribution of the gauge fields to $\gamma_{ij}$ is taken into account. 

{\it Linear perturbation and polarization bases.}--In the conventional $\delta N$ formalism, the adiabatic curvature perturbation $\zeta$ is given by computing the $e$-folding number under different initial conditions. Similarly, we can compute GWs simply by computing the anisotropic expansion, which is given by Eq.~(\ref{Sol:gamma}). In this letter, we consider the linear perturbation except for the paragraph of the primordial non-Gaussianity.

In what follows, we set the background spatial metric at the reheating surface $t_f$ as $\bar{\gamma}_{ij}(t_f) = \delta_{ij}$. Then, at $t=t_f$, we can define GWs as usual by using the polarization tensors $e_{ij}^{(\lambda_{\rm gw})}$ with $\lambda_{\rm gw}= +,\, \times$, which satisfy $  \hat{k}_i e_{ij}^{(\lambda_{\rm gw})}  (\hat{\bbm{k}})= 0$, $e_{ii}^{(\lambda_{\rm gw})}  (\hat{\bbm{k}})= 0$, and $ e_{ij}^{(\lambda_{\rm gw})}  (\hat{\bbm{k}}) e_{ij}^{(\lambda'_{\rm gw})}  (\hat{\bbm{k}}) = \delta^{\lambda_{\rm gw} \lambda'_{\rm gw}}$. The adiabatic curvature perturbation is also given by the usual definition as $\zeta(t_f) = \delta \psi(t_f) - \frac{1}{4} \hat{k}_i \hat{k}_j \delta \gamma_{ij}(t_f)$. Here, $\hat{\bbm{k}}$ denotes the unit wavenumber $\bbm{k}/k$. Here and hereafter, we lower and raise the indices $\hat{k}^i$ and $e_{ij}^{(\lambda_{\rm gw})}$ by using $\bar{\gamma}_{ij}(t_f) = \delta_{ij}$ and $\bar{\gamma}^{ij}(t_f) = \delta^{ij}$. (When the background shear had already become negligible at $t=t_f$, the background spatial metric remains  $\bar{\gamma}_{ij}(t) = \delta_{ij}$ all the time after $t_f$, ensuring the linear decomposition among the scalar, vector, and tensor type perturbations.) Using the two polarization bases of the gauge fields, $e^{(\lambda)}_i(\hat{\bbm{k}})$ with $\lambda=1,\, 2$, we define $e_{ij}^{(\lambda_{\rm gw})}(\hat{\bbm{k}})$ as $e^{(+)}_{ij} \equiv (e_i^{(1)} e_j^{(1)} -e_i^{(2)} e_j^{(2)})/\sqrt{2}$ and $e^{(\times)}_{ij} \equiv (e_i^{(1)} e_j^{(2)} +e_i^{(2)} e_j^{(1)})/\sqrt{2}$.

For our later use, let us introduce the polar and azimuthal angles as $ \hat{{\bbm k}} \cdot \bar{\hat{\bbm{\pi}}}_{(\alpha)}= \cos \Theta_\alpha$, $ \bbm{e}^{(1)} \cdot \bar{\hat{\bbm{\pi}}}_{(\alpha)} = \sin \Theta_\alpha  \cos \Psi_\alpha$, and $\bbm{e}^{(2)} \cdot \bar{\hat{\bbm{\pi}}}_{(\alpha)} = \sin \Theta_\alpha  \sin \Psi_\alpha$ for each $\alpha$, where $\bar{\hat{\pi}}_{(\alpha)}^i$ denotes the background component of the normalized conjugate momenta, given by
\begin{align}
    \hat{\pi}_{(\alpha)}^i \equiv \frac{\pi_{(\alpha)}^i}{\sqrt{\gamma_{ij*} \pi_{(\alpha)}^i \pi_{(\alpha)}^j}} = \frac{\gamma^{ij}_* \dot{A}_{j(\alpha)*}}{\sqrt{\gamma^{kl}_* \dot{A}_{k(\alpha)*} \dot{A}_{l(\alpha)*}}}\,. 
\end{align}
For an arbitrary $\hat{k}^i$, we can choose $e_i^{(2)}$ such that being orthogonal to one of the background gauge fields, e.g., $\Psi_1= 0$. In fact, we can choose $e_i^{(1)}$ in the 2D plane spanned by $\hat{k}_i$ and $\bar{\hat{\pi}}_{(1)}^i$, ensuring $e_i^{(2)} \bar{\hat{\pi}}^i_{(1)}=0$.

 {\it Linearly polarized GWs.}--The two linear polarization modes, defined as $ \gamma^{(\lambda_{\rm gw})}(t_f,\, \bbm{k}) \equiv e_{ij}^{(\lambda_{\rm gw})}(\hat{\bbm{k}})\, \delta \gamma_{ij} (t_f,\, \bbm{k})$, in general evolve differently in superhorizon scales, generating the linear polarization. Here, let us discuss when the linear polarization becomes non-zero, focusing on the case where the two linear polarization modes of the gauge fields have the same amplitudes at $t_*$, i.e., $|e_i^{(1)} \delta \pi_{(\alpha)}^i|= |e_i^{(2)} \delta \pi_{(\alpha)}^i|$ (without having cross-correlation among them). When the fluctuations of ${\cal I}_{(\alpha)f}$ and $\gamma_{ij*}$ are negligible in the second term of Eq.~(\ref{Sol:gamma}), i.e., when the fluctuation of the gauge field $\delta \pi^i_{(\alpha)}$ yields the dominant contribution, this sourced contribution does not generate the linear polarization. Since the contribution of $\delta \gamma_{ij*}$ becomes smaller than that of $\delta \pi^i$ as will be discussed shortly, let us focus on the fluctuation of ${\cal I}_{(\alpha)f}$. Since ${\cal I}_{(\alpha)f}$ is the time integral of the functional of the scalar fields, $\delta {\cal I}_{(\alpha)f}$ is generated by the intrinsic fluctuations of the scalar fields and the fluctuations sourced by the gauge fields. Even if the intrinsic one is negligible, when the backreaction of the increasing gauge fields on the scalar fields starts to be important, the fluctuations of the gauge fields perturb ${\cal I}_{(\alpha)f}$ through the sourced fluctuations, resulting in the superhorizon generation of the linear polarization. In the model of anisotropic inflation discussed in Refs.~\cite{Watanabe:2009ct, Watanabe:2010fh} (exclusively for the specific parameter choice there), the gauge field does not effectively change the equation of motion of the scalar field. Therefore, the linear polarization was negligibly small \cite{Tanaka:2023}.

{\it Slow-roll evolution and asymptotic solution.}-- Let us introduce $\hat{{\cal I}}_{(\alpha)f} \equiv\gamma_{ij*} \pi_{(\alpha)}^i \pi_{(\alpha)}^j{\cal I}_{(\alpha)f}$, which is given by 
\begin{align}
    & \hat{{\cal I}}_{(\alpha)f}(\{\varphi^{a'}_*\}') \simeq \frac{2}{\rho_{*}} \int^{\psi_f}_{\psi_*}  d \psi \rho_{A(\alpha)} (\psi),  \label{Def:calI} 
\end{align}
since the time variation of $\gamma_{ij}$ becomes higher order and $\pi^i_{(\alpha)}$ is constant in time. When $\rho_{A(\alpha)}$ takes a maximum value $\rho_{A(\alpha)}^{\rm max}$ during $\Delta \psi_{(\alpha)}$ after the horizon crossing, $\If$ is roughly given by $\If \sim 2 \Delta \psi_{(\alpha)} \rho_{A(\alpha)}^{\rm max}/\rho_*$. When the scalar fields which directly interact with the gauge fields undergo the slow-roll evolution, one can broadly find a solution where $\rho_{A(\alpha)}$ approaches an almost constant value, whose time variation is suppressed by the slow-roll parameters~\cite{Tanaka:2023}. An explicit example can be found e.g. in Ref.~\cite{Fujita:2018zbr}. In such cases, 
\begin{align}
  ({\rm A2}) \quad \left| \frac{\delta \If}{\If} \right| \ll |\delta \hat{\pi}_{(\alpha)}^i|\,
\end{align}
naturally holds, because the fluctuation of $\rho_{A(\alpha)}^{\rm max}$ is suppressed and $|\delta \Delta \psi_{(\alpha)}|/\Delta \psi_{(\alpha)}$ typically becomes smaller than $|\delta \hat{\pi}^i_{(\alpha)}| \! \sim \! |\delta \rho_{A(\alpha)*}/\rho_{A(\alpha)*}|$ by a factor of $1/\Delta \psi_{(\alpha)}$. In order for the condition (A2) to hold, the fluctuations of the scalar fields and the amplitude of each gauge field, $\Bar{\gamma}_{ij} \Bar{\pi}^i_{(\alpha)} \delta \pi^j_{(\alpha)}$, need to be tightly correlated to reduce the fluctuation of $\rho_{A(\alpha)}$. In this case the backreaction of the gauge field modifies the evolution of the scalar fields, resulting in the non-zero linear polarization (see Eq.~(\ref{Def:gt})). In the anisotropic inflation model considered in Refs.~\cite{Watanabe:2009ct, Watanabe:2010fh}, (A2) does not hold exceptionally for the parameter chosen there (while GWs can be still computed by using Eq.~(\ref{Sol:gamma})).

{\it Perturbative expansion in \dNex.}-- Perturbing Eq.~(\ref{Sol:gamma}) under the assumption (A2), the linear perturbation of $\gamma_{ij}$ is given by (see Supplementary material)
\begin{align}
   & \delta \gamma_{ij} (t_f,\, \bbm{x})  \simeq \delta \gamma_{ij*}(\bbm{x})     -  4 \left[ \bar{\gamma}_{il*} \bar{\gamma}_{jm*}\right]^{\rm TL} \hspace{-2pt} \bar{\hat{\pi}}^{\{l}_{(\alpha)} \delta \hat{\pi}^{m\}}_{(\alpha)}(\bbm{x})  \bar{\hat{{\cal I}}}_{(\alpha)f},  \label{Sol:gamma2}
\end{align}
where the indices $l$ and $m$ are symmetrized. The spatial inhomogeneity at $t=t_*$ is described by considering separate universes with various initial conditions~\cite{Tanaka:2021dww}. The first term is the usual vacuum contribution, while the second one describes the shear fluctuation sourced by the directional fluctuations of $\pi^i_{(\alpha)}$. The fluctuation of $\gamma_{ij*}$ also perturbs the second term of Eq.~(\ref{Sol:gamma}), yielding several terms whose amplitudes amount to 
$\bar{\hat{{\cal I}}}_{(\alpha)f} (H_*/\Mp) \sim (\bar{\rho}^{\rm max}_{A(\alpha)}/\bar{\rho}_*) \Delta \psi_{(\alpha)} (H_*/\Mp)$, where $H$ denotes the Hubble parameter. They turn out to be smaller than the second term of Eq.~(\ref{Sol:gamma2}) by $\sqrt{\bar{\rho}^{\rm max}_{A(\alpha)}/\bar{\rho}_*} < 1$. 
This suppression originates from the difference between the amplitudes of $\delta \gamma_{ij*}$ and $\delta \hat{\pi}^i_{(\alpha)}$. A more general computation in which (A2) is not imposed will be reported in Ref.~\cite{Tanaka:2023}.

Perturbing $\hat{\pi}^i_{(\alpha)}$, we obtain 
\begin{align*}
   \delta \hat{\pi}_{(\alpha)}^i  = \frac{\bar{\gamma}^{ij}_* \delta \dot{A}_{j(\alpha)*}}{\dot{\bar{A}}_{(\alpha)*}} - \frac{\bar{\gamma}_*^{kl} \dot{\bar{A}}_{k(\alpha)*} \delta \dot{A}_{l(\alpha)*} \bar{\gamma}^{ij}_* \dot{\bar{A}}_{j(\alpha)*} }{\dot{\bar{A}}_{(\alpha)*}^3} + \cdots\,, 
\end{align*}
where we have introduced $\dot{\bar{A}}_{(\alpha)*} \equiv \sqrt{\bar{\gamma}^{kl}_* \dot{\bar{A}}_{k(\alpha)*} \dot{\bar{A}}_{l(\alpha)*}}$. Here we abbreviated the terms with the fluctuation of the spatial metric, which only give sub-dominant contributions. Without these abbreviated terms, we find $\bar{\gamma}_{ij*} \delta \hat{\pi}_{(\alpha)}^i \bar{\hat{\pi}}^j_{(\alpha)}=0$.

{\it Power spectrum of GWs.}-- So far, we have computed the mapping between the horizon crossing $t_*$ and the reheating $t_f$ simply by assuming (A1) and (A2). In what follows, assuming further that the background anisotropy was still very small at $t=t_*$, we compute the power spectrums of $\delta \gamma_{ij*}$ and $\delta \dot{A}_{i*}$ by adopting the FLRW background approximation. The amplitudes of $\delta \dot{A}_{(\alpha)*}$ for the two polarization modes are given by 
\begin{align*}
    \frac{|\delta \dot{A}_{(\alpha)*} (k)|^2}{\dot{\bar{A}}_{(\alpha)*}^2}  = \frac{1}{12}\! \left( 1 + \left( \frac{\dot{\bar{f}}_{*}}{H_{*}\Bar{f}_*} \right)^{\!\!2} \right)\! \frac{1}{k^3} \frac{\Bar{\rho}_{*}}{\Bar{\rho}_{A (\alpha) *}} \left( \frac{H_{*}}{\Mp} \right)^2\hspace{-5pt},
\end{align*}
which should be smaller than 1 to validate the perturbation, resulting in the upper bound on the duration of the exponential growth of each gauge field. Combining the expressions given above, we obtain the power spectrum of primordial GWs as
\begin{align}
    &\langle \gamma^{(\lambda_{\rm gw})} (t_f,\, \bbm{k}) \gamma^{(\lambda_{\rm gw})} (t_f,\, \bbm{p}) \rangle \nonumber \\
    &= \delta (\bbm{k} + \bbm{p}) \frac{2}{k^3} \left( \frac{H_{*}}{\Mp} \right)^{\!\!2}\!  \left( 1 +  \sum_{\alpha=1}^{D'} g_{t (\alpha)}^{\lambda_{\rm gw}} \sin^2 \Theta_\alpha \right) \,, \label{Def:gt}
\end{align}
where $g_{t (\alpha)}^{\lambda_{\rm gw}}$ depends on the angles $\Theta_\alpha$ and $\Psi_\alpha$ as  
\begin{align*}
    & g_{t(\alpha)}^{+} = g_{t(\alpha)} \cos 2 \Psi_\alpha \! \left(1 - 2  \sin^2 \Theta_\alpha +  \sin^4 \Theta_\alpha \cos^2 2 \Psi_\alpha \! \right)\!\!,\\
    & g_{t(\alpha)}^{\times} =g_{t(\alpha)} \cos 2 \Psi_\alpha \! \left(1 - \sin^4 \Theta_\alpha \sin^2 2\Psi_\alpha \! \right)\!\!, 
\end{align*}
with the overall amplitude $g_{t(\alpha)}$ being
\begin{align}
    & g_{t(\alpha)} \equiv  \frac{\Bar{\rho}_*}{3 \Bar{\rho}_{A(\alpha)*}} \bar{\hat{{\cal I}}}_{(\alpha)f}^2 
    \left( 1 + \left( \frac{\dot{\bar{f}}_{(\alpha)*}}{H_{*}\Bar{f}_{(\alpha)*}} \right)^{\!\!2} \right), 
\end{align}
which is ${\cal O} ( (\bar{\rho}_{A(\alpha)}^{\rm max}/\bar{\rho}_*) (\bar{\rho}_{A(\alpha)}^{\rm max}/\bar{\rho}_{A(\alpha)*}) \Delta \psi_{(\alpha)}^2)$. This indicates that even if the energy density of the gauge field all the time remains much smaller than the total energy density, satisfying $(\bar{\rho}_{A(\alpha)}^{\rm max}/\bar{\rho}_*) \Delta \psi_{(\alpha)} ^2 \ll 1$, the anisotropic component of primordial GWs can be as large as or even larger than the isotropic one \footnote{As long as $ (\rho_A^{\rm max}/\rho_*) \Delta \psi$ is smaller than 1, we can still neglect the higher order terms with respect to $\rho_A/\rho$ which are ignored e.g. in (\ref{Sol:gamma})}. This is a consequence of the enhancement by $\bar{\rho}_{A(\alpha)}^{\rm max}/\bar{\rho}_{A(\alpha)*}$, which becomes much larger than 1 when the mode $k$ crosses the horizon before $\rho_{A(\alpha)}$ reaches the maximum value. This is because the conversion from the fluctuations of the gauge fields to GWs takes place mainly when $\rho_{A(\alpha)}$ reaches the maximum value, while the (normalized) power spectrum of the gauge field at the horizon crossing is inversely proportional to $\Bar{\rho}_{A(\alpha)*}$. Because of this enhancement, even if the gauge field is only sourced by the spectator fields, which only occupy a small fraction of the total energy density, we can obtain a large statistical anisotropy~\cite{Fujita:2018zbr}. Since the backreaction is important when (A2) holds as discussed before, the two linear polarization modes have a different angular dependence. In general, the sourced GWs depend both on the polar and azimuthal angles, but exclusively for $D'=1$, we can eliminate the latter by setting it e.g. to $\pi$. The cross-correlation between $\gamma^{(+)}$ and $\gamma^{(\times)}$ necessarily takes a non-vanishing value for $D' \geq 2$, since it turns out to be proportional to $\sin \Psi_\alpha$, which cannot be set to 0 for all $\alpha$s simultaneously \cite{Tanaka:2023}.

{\it Adiabatic curvature perturbation.}-- The (linear) adiabatic perturbation is given by the summation of the fluctuation of the $e$-folding number and the longitudinal part of $\delta \gamma_{ij}$ at $t=t_f$, where the background spatial metric is set to $\delta_{ij}$. Under the assumptions (A1) and (A2), we obtain the dominant contribution of the latter as
\begin{align*}
    \frac{1}{4} \hat{k}_i \hat{k}_j \delta \gamma_{ij}(t_f,\, \bbm{k}) = \sum_{\alpha=1}^{D'} (\hat{k}_i \bar{\hat{\pi}}^i_{(\alpha)})^2  \bar{\hat{{\cal I}}}_{(\alpha)f} \bar{\hat{\pi}}^j_{(\alpha)} \frac{\delta \dot{A}_{j*(\alpha)}(\bbm{k})}{\dot{\bar{A}}_{(\alpha)*}}\,,
\end{align*}
while the former depends significantly on the detail of the models. Here, we have dropped the longitudinal mode of $\delta \gamma_{ij}$ at $t=t_*$, which can be removed by using the residual gauge degree of freedom. 

For example, when there exist two canonical scalar fields $\phi$ and $\sigma$ and only the subdominant one $\sigma$ interacts with one gauge field $(i.e., D=2, D'=1)$ under the slow-roll approximation, we obtain
\begin{align*}
    \frac{d \phi}{d \psi} \simeq - \Mp^2  \frac{V_\phi}{V + \rho_{\sigma} + \rho_A}\,, 
\end{align*}
where $V$ and $V_\phi$ denote the scalar potential and its derivative with resect to $\phi$ and $\rho_\sigma$ denotes the energy density of $\sigma$. Assuming that when the energy density of the gauge field takes the maximum value $(\psi_{\rm max} \leq \psi \leq \psi_{\rm max} + \Delta \psi )$ both $\rho_A$ and $\rho_{\sigma}$ remain almost constant, the above equation can be solved as
\begin{align}
   \psi_f - \psi_* \simeq  (\psi_f - \psi_*)_{\phi} + \frac{\rho^{\rm max}_{\sigma} + \rho^{\rm max}_A}{V} \Delta \psi\,,  \label{Eq:efolding}
\end{align}
where the first term denotes the $e$-folding number, which is determined only by the dynamics of $\phi$. Here and hereafter, we put the upper index max on the quantities evaluated during $\psi_{\rm max} \leq \psi \leq \psi_{\rm max} + \Delta \psi $. The subdominant scalar field $\sigma$ can also provide an angular-independent sub-dominant contribution to $\psi_{\rm max} - \psi_*$, which can be addressed by using the conventional $\delta N$ formalism. Here and hereafter, we ignore them, focusing on the leading angular dependent contribution. Perturbing Eq.~(\ref{Eq:efolding}), we find that the dominant angular dependent contribution appears from the fluctuation of $\rho_\sigma^{\rm max}$ under the assumption (A2). Using $\delta \rho_\sigma^{\rm max} \simeq \bar{V}^{\rm max}_{\sigma} \delta \sigma^{\rm max}$ and eliminating $\delta \sigma^{\rm max}$ by using $\delta \rho^{\rm max}_{A} \simeq 0$, we obtain
\begin{align*}
   \delta \rho_{\sigma}^{\rm max} \simeq \frac{\bar{V}_{\sigma}^{\rm max}}{(\ln \bar{f})_{, \sigma^{\rm max}}} \left( \Bar{\hat{\pi}}^i \frac{\delta \dot{A}_{i*}}{\dot{\bar{A}}_*}  + 2 (\ln \Bar{f})_{, \sigma *} \delta \sigma_* \right)\,.  
\end{align*}
Using this in the perturbation of Eq.~(\ref{Eq:efolding}) and ignoring the second term in $\delta \rho_{\sigma}^{\rm max}$, which only yields the subdominant angular independent contribution, we obtain
\begin{align*}
    \zeta(t_f,\, \bbm{k}) &\simeq - \frac{\bar{V}_*}{\Mp^2 \bar{V}_{\phi*}} \delta \phi_*(\bbm{k})  - 2 \frac{\bar{\rho}_A^{\rm max}}{\bar{\rho}_*} \Delta \psi \frac{\bar{\hat{\pi}}^i \delta \dot{A}_{i*}(\bbm{k})}{\dot{\bar{A}}_*} \cr
    & \qquad \qquad  \times \left(  \cos^2 \Theta(\hat{\bm{k}}) - \frac{1}{2} \frac{V_{\sigma}^{\rm max}}{ (\ln f)_{, \sigma^{\rm max}} \Bar{\rho}_{A}^{\rm max}} \right) \,.
\end{align*}
The first term in the parenthesis is the model independent contribution which comes from the longitudinal mode of $\delta \gamma_{ij}$ and the second one comes from the fluctuation of the $e$-folding number. For the specific choice of $f$ and $V$, the formulae given in this letter reproduce the result obtained in Ref.~\cite{Fujita:2018zbr}. When the amplitude of the second term in the parenthesis of $\zeta$ does not exceed ${\cal O}(1)$, the statistical anisotropy in the power spectrum of $\zeta$ is suppressed by the slow-roll parameter $\varepsilon \simeq (\Mp V_\phi/V)^2/2$ compared to that of GWs $g_t^{\lambda_{\rm gw}}$. Then, the statistical anisotropy in the power spectrum of GWs can be as large as ${\cal O}(1)$, keeping the scalar spectrum consistent with the present observations~\cite{Fujita:2018zbr}. Using these formulae, we can also compute the cross-correlation of $\zeta$ and $\gamma^+$, which takes a non-vanishing value.

{\it Primordial non-Gaussianity.}-- Using the \dNex, the local type non-Gaussianity can be computed easily. Here, let us only provide the order estimation, leaving a detailed computation for elsewhere. We obtain the local type non-Gaussianity of $\gamma^{(\lambda_{\rm gw})}$ sourced by the gauge fields as 
\begin{align*}
    f_{\rm NL}^{\gamma, A} \sim \sum_{\alpha=1}^{D'} \Bigl( \frac{\bar{\rho}_*}{\bar{\rho}_{A(\alpha) *}} \Bigr)^{\!\!2} \, \bar{\hat{{\cal I}}}_{(\alpha)f}^3 \sim \sum_{\alpha=1}^{D'} \Bigl( \frac{\bar{\rho}_{A(\alpha)}^{\rm max}}{\bar{\rho}_{A(\alpha) *}} \Bigr)^{\!\!2} \frac{\bar{\rho}_{A(\alpha)}^{\rm max}}{\bar{\rho}_*} \Delta \psi_{(\alpha)}^3 \,,
\end{align*}
which can be also enhanced by the square of $\bar{\rho}_{A(\alpha)}^{\rm max}/\bar{\rho}_{A(\alpha) *}$, when the corresponding gauge field reaches the maximum value after the horizon crossing. Here, we have ignored the angular dependence which requires a more detailed computation. The local type non-Gaussianity of $\zeta$ which stems from the longitudinal part of $\delta \gamma_{ij}$ is suppressed by $\varepsilon^2$ compared to $f_{\rm NL}^{\gamma, A}$, while there is also model dependent contribution in the fluctuation of the $e$-folding number.

{\it Summary.}-- In this letter, we showed the \dNex~can largely facilitate the computation of the superhorizon evolution of $\zeta$ and GWs by considering an inflation model with U(1) gauge fields. Since the \dNex~generically applies to a model with the locality and the spatial diffeomorphism invariance, various applications will be possible. The \dNex~plays the complementary role to the cosmological bootstrap~\cite{Arkani-Hamed:2018kmz}, which is powerful to address the correlation functions at the horizon crossing.   

The \dNex~is useful to develop the intuitive understanding about the large scale evolution. Along this line, we discussed the condition for the nono-zero linear polarization and showed the qualitative difference between single- and multi-gauge field models as an example. 

The model studied here has various phenomenologically interesting properties, e.g., predicting the $O(1)$ statistical anisotropy in the power spectrums of GWs, the linear polarization, and the cross-correlation among $\zeta$ and the two linear polarization modes of GWs, which will leave characteristic signals e.g., in the fluctuations of the cosmic microwave background~\cite{Watanabe:2010cmb}.

\begin{acknowledgments}
T.~T. is supported by Grant-in-Aid for Scientific Research (A) and (C) under Contract Nos. JP23H00110 and JP20K03928, respectively. Y.~U. is supported by Grant-in-Aid for Scientific Research (B) under Contract Nos.~JP19H01894 and JP23H01177 and Fostering Joint International Research (B) under Contract No.~JP21KK0050, and JST FOREST Program under Contract No.~JPMJFR222Y.
\end{acknowledgments}

\bibliography{refst}

\end{document}